\documentclass[acmsmall]{acmart}

\AtBeginDocument{%
  \providecommand\BibTeX{{%
    \normalfont B\kern-0.5em{\scshape i\kern-0.25em b}\kern-0.8em\TeX}}}

\usepackage{arabtex}
\usepackage{graphicx}
\usepackage{tabularx}
\usepackage{booktabs}
\usepackage{caption}
\usepackage{subcaption}
\usepackage{graphicx}
\usepackage{multicol}
\usepackage{graphicx}
\usepackage{comment}

\usepackage[ruled]{algorithm2e}

\usepackage{xcolor,pifont}
\newcommand*\colourcheck[1]{%
 \expandafter\newcommand\csname #1check\endcsname{\textcolor{#1}{\ding{52}}}%
}
\colourcheck{blue}
\colourcheck{green}
\colourcheck{red}

\usepackage{utf8}
\setcode{utf8}

\begin{document}

\title{Holy Tweets: Exploring the Sharing of Quran on Twitter}

\author{Norah Abokhodair}
\affiliation{
  \institution{Microsoft}
  \city{Redmond}
  \country{USA}}
\email{noraha@uw.edu}
\author{Abdelrahim Elmadany}
\affiliation{%
  \institution{University of British Columbia}
  \city{Vancouver}
  \country{Canada}}
\email{a.elmadany@ubc.ca}
\author{Walid Magdy}
\affiliation{%
  \institution{School of Informatics, The University of Edinburgh}
  \city{Edinburgh}
  \country{UK}}
\email{wmagdy@inf.ed.ac.uk}

\renewcommand{\shortauthors}{}


\begin{abstract}
While social media offer users a platform for self-expression, identity exploration, and community management, among other functions, they also offer space for religious practice and expression.  In this paper, we explore social media spaces as they subtend new forms of religious experiences and rituals. We present a mixed-method study to understand the practice of sharing Quran verses on Arabic Twitter in their cultural context by combining a quantitative analysis of the most shared Quran verses, the topics covered by these verses, and the modalities of sharing, with a qualitative study of users' goals. This analysis of a set of 2.6 million tweets containing Quran verses demonstrates that online religious expression in the form of sharing Quran verses both extends offline religious life and supports new forms of religious expression including goals such as doing good deeds, giving charity, holding memorials, and showing solidarity. By analysing the responses on a survey, we found that our Arab Muslim respondents conceptualize social media platforms as everlasting, at least beyond their lifetimes, where they consider them to be effective for certain religious practices, such as reciting Quran, supplication (dua), and ceaseless charity. Our quantitative analysis of the most shared verses of the Quran underlines this commitment to religious expression as an act of worship, highlighting topics such as the hereafter, God's mercy, and sharia law. We note that verses on topics such as jihad are shared much less often, contradicting some media representation of Muslim social media use and practice. 
\end{abstract}

\maketitle
\section{Introduction}
\label{sec:intro}

Religious and spiritual rituals are important aspects of Muslim societies and are interwoven in the fabric of everyday life. In the Arab-Muslim society, Islamic values and practices are tangibly, visibly, and audibly built into the environment: the call to prayer that one hears five times a day, the daytime closing of restaurants during the fasting hours of Ramadan, and the distribution of mosques throughout neighborhoods to support the value of praying on time. They are also embodied in the unspoken guiding principles for familial and non-familial relationships, marriages, and hospitality, amongst other practices. Whereas social media studies to date has placed a disproportionate focus on western users' online expression, here we turn to the Arab world to examine religious uses of social media in their situated cultural and geographic contexts.

In today's networked and tech-infused era, these practices are increasingly taking nontraditional shapes and forms. With the pervasive use of technology and social media in the Arab World, reaching over 15 millions Twitter users in Saudi Arabia alone\cite{clement_2020}, a wave of practices involving new interpretations and meanings to some prominent Islamic rituals and practices are on the rise with no one to trace them back to their theological origin nor explain how and why they emerged. Technologies such as smart phones and tablets are influencing many Muslim rituals given the increased accessibility and development that resulted in increasing numbers of applications that support and mediate a variety of Muslim everyday rituals. Searching the Apple App Store or Google Play Store for terms like "Muslim" returns results for applications to point users to Mecca, inform them of local prayer times, provide digital versions of Quran, and a mosque-function that prevents your phone from ringing during prayer~\cite{OnePath_2019}. These technologies are also offering means to digitizing many of the prominent religious artifacts. Take, for example, a scene from Al-Haram Mosque (i.e. the Great Mosque of Mecca) during Hajj, the annual pilgrimage required once in a Muslim's lifetime. Large numbers of people now bring mobile phones or tablet devices to recite the Quran, remember prayer times, or record dua in the place of traditional prayer beads or scripture.

Scholars of social media to date have largely engaged religious communities from a social networking perspective to study how they utilize, engage with, and appropriate technology including social media in their everyday life. This growing body of literature had brought our attention to many emerging themes regarding self-presentation and expression, identity construction, and communicating and socializing with others (e.g., \cite{marwick2011tweet,madden2013teens}). However, little consideration is afforded to the religious facet of users' lives, for example, how this technology supports their religious rituals and the content or purpose of what is being shared. This under-representation has been historically noticed; the seminal research of both Genevieve Bell~\cite{bell2006no} on \textit{techno-spiritual practices} and Heidi Campbell ~\cite{campbell2012digital} in \textit{Digital Religion} warns of the consequences of neglecting the religious dimension of users' contemporary lives because it can lead to misinterpreting important social phenomena. 

Inspired by the seminal research of Bell and Campbell\cite{bell2006no, campbell2012digital}, our current study dives into the ways Islamic rituals are transcending offline boundaries and finding new forms of expression online through studying Quran sharing on Arabic Twitter. Specifically, we share findings from a mixed-method study that looks into the depths of the means and characteristics of Quran, the central religious text of Islam, share on Twitter.

 Outside of CSCW, popular media representations of Islamic rituals in research, especially Quran-related ones, have been distorted. 
Multiple studies examine the use of Quran verses in large datasets for the detection of radicalism and the identification of Jihadists recruiting by ISIS and other terrorist groups on social media, e.g., \cite {wright2016resurgent, macdonald2019daesh,cain_gonzalez_2017}. Other studies focus on the extreme religious interpretations of the Quran and its relationship to violence and radicalism, e.g., \cite{venkatraman2007religious,leezenberg2007islamic}. Our research sought to understand how Arab Muslims enact Quran-related rituals online. We do that by analyzing a large Arabic Twitter dataset that includes Quran verses (2.6 million tweets collected over two years), together with the results of a qualitative survey study (n=135) and triangulated these with 7 in-depth interviews. In this paper, we pay special attention to how Quran verses are being shared, which ones, and by whom. In taking this approach in studying religious online expression, grounded in understanding the practice of Quran-related rituals and how they manifest online, we enrich the scholarly understanding of the use of Quran verses online. We argue that studying religious expression online is a valuable means to understand social media's influence on society and culture.

\subsection{Motivation}
For the Muslim population, research supports that many online practices are influenced by Islamic teachings and beliefs ~\cite{abokhodair2017transnational,abokhodair2017photo,alshehri2018beauty,Nassir:2019:MGD:3328418.3318145,10.1145/3359148,10.1145/3359148,10.1145/3025453.3025960,Alsheikh:2011:VDS:1958824.1958836,Vieweg:2016:SMS:2818048.2819966,al2017against}. We assert that interpreting users' behavior from a holistic lens, i.e., including their religious motivations, leads to a deeper understanding of this user population and and goes hand-in-hand to explain online decisions and behaviors~\cite{dawson2013religion,campbell2010religion}. Whether online or offline, understanding religious values and rituals for this population in the context of CSCW is no longer an optional goal, but core for developing empathy- and inclusion-based technologies. Through analysing Arabic Quran content sharing, we engage this material on its own terms, widening the boundaries of CSCW to embrace an Arab religious cultural context while providing an example of how technology is transforming and mediating Muslim worship, the central discussion in this paper.

 Since social media has become the space for practicing many offline religious rituals, the CSCW community is an appropriate home to longitudinal scholarship on what was and is technology mediated spiritual practice~\cite{10.1145/3290607.3310426}. We invite the CSCW community to view Arab users' online behaviours holistically, within their offline cultural context, in order to understand Arab CSCW. This ongoing research contributes to the collective effort of documenting and archiving the longitudinal narrative through academic publishing, to clarify the bi-directional effect of technology and religion.

Finally, we undertake this research in recognition of the growing practice of sharing the Quran and other religious symbolism on Twitter and other social media platforms, and the sometimes erroneous explanation for these practices offered by public media, which can potentially harm or disadvantage Muslim users. 

\subsection{The study \& Research Questions}

 The broad questions we ask are: R1) What role does online expression play in religious life for Muslims from the Arab world? R2) How do the affordances of social media influence religious expression online? R3) How do the affordances of social media foster emergent religious practices? The sub-research questions we tackle in this paper are: 1) How do Arab-Muslims share the Quran on Twitter and why? 2) What are the most shared Quran verses on Twitter? 3) What are they about? and by whom?

For this exploratory analysis, we draw on data we collected iteratively through a mixed-methods study. We began by collecting Arabic tweets containing Quran verses over a 2 year period resulting in a dataset of 2.6 million tweets that were retweeted over 9 million times. We then conducted an automated content analysis on the tweets collection to categorize them in Quran-specific topics. As new findings emerged from our quantitative analysis, we decided to further our inquiry through instrumenting a qualitative survey and administrating it on Twitter for 3 months (n=135). Finally, we conducted in-depth interviews with seven Arab-Muslim participants to triangulate and further develop observations emergent in the survey. 

This study makes the following contributions. It characterizes the topics covered in the Quran versus and explored what verses are shared on Twitter and the purposes for which these verses are shared. During our study, we observed the use of automation to support users' sharing verses from Quran, and offer reflections on the rationales for adopting automated agents as a form of technology-mediated worship. Furthermore, we discover that Twitter, as a platform is perceived as everlasting, or at least longer than a user's lifetime, which motivates them to use it for Muslim rituals such as ongoing charity and digital memorials. Our dataset encompasses tweets over the course of the tragic Christchurch mosque shooting in New Zealand. We further develop our analysis on the role of online religious expression during that devastating event.

\section{Background}
To situate our research in its larger context and help readers better understand the concepts presented in this paper,
we provide a brief background on some of the dominant aspects of life in the Arab World, focusing on Islam, its relevant practices, and the Quran, the textual focus of this study.

\subsection{Islam and The Arab World}
The Arab world consists of 22 countries stretched across two contents (Asia and Africa), making it one of the world's most strategic territories. It is rich in diversity with many religious, ethnic, and linguistic groups sharing the same region. Arabic is the official language and the language of the Quran. In terms of the Islamic faith (both within and outside of the Arab world), a recent report estimated that there were 1.8 billion Muslims in the world as of 2015, roughly 24\% of the global population~\cite{lipka_2017}, making Islam the world's second-largest religious tradition after Christianity. Arab Muslims, the focus of this study, constitute 20\% of the world's Muslim population~\cite{masci_2017}.

\subsection{The Muslims' Holy Book: The Quran}
In Muslim cultures, cultural practices pertaining to morality and ethics are primarily derived from Quranic text, and the quotes of the prophet Muhammad (i.e. Hadith). Muslims believe that the Quran was revealed to the Prophet Muhammad in 610 CE~\cite{haleem2010qur,nasr_2017}, and along with authentic Hadith is the main source for the interpretation of the often-practiced law in Muslim countries (i.e., sharia law). The Quran is divided into 114 chapters (called \textit{suras} in Arabic), with 6,236 total verses (known as \textit{ayat} in Arabic)~\cite{haleem2010qur,nasr_2017}.
The content of the Quran describes the basic Islamic beliefs, including general exhortations regarding right and wrong, narratives of the early prophets and old nations, Judgement Day and the afterlife, and other topics. The Quran emphasizes the importance of preparing for the afterlife, and the accountability for one's actions, whether good or bad, caused by the accumulation of good deeds or sins. 
Relevant to this study, good deeds are considered to follow a person through eternity; the death of a person does not mean the death of the actions they took while alive. For example, printing a Quran and leaving it in a mosque for others to read is believed to be a source of good deeds, whether the person who did so is alive or deceased~\cite{kauthar_2017}. This practice is referred to as \textit{ceaseless charity} (\textit{sadaqah jariyah} in Arabic), defined as "having a belief in continuous charity by the mode of good deeds and action''~\cite{kauthar_2017}, that will not only benefit one in this life, but will continue to benefit one after their death. According to a Hadith, where ceaseless charity is referenced: ``When a person dies, their deeds come to an end except for three: ceaseless charity, beneficial knowledge, or a righteous child praying for them''. 
In addition, Muslim belief puts importance on the reciting and sharing, orally or in writing, verses from the Quran. If a Quran verse is shared and repeated by another person, then both the reciter and repeater are rewarded with the good deed and it is further multiplied by more sharing~\cite{kauthar_2017}.

\section{Related work}
 Before we delve into the details of our current study, we provide a brief overview of relevant CSCW, HCI, and digital religion research in two areas. First, we review the literature on technology-mediated worship and spirituality. Second, we discuss the studies that analyze religious expression online with a focus on two of the most popular social media platforms in the Arab world: Twitter and Instagram. Our work intersects with and extends these areas in key ways, specifically by considering religious ritual application, expression, and mediation in new media in a non-Western Islamic context.

\subsection{Technology mediated worship and spirituality (Techno-spirituality)}
In reviewing relevant research, we engage with general HCI and CSCW work on the use of technology that supports religious activities, rituals and literature on techno-spiritual practices ~\cite{bell2006no,Woodruff:2007:SDH:1240624.1240710,Wyche:2008:SDE:1358628.1358866,Wyche:2006:TSF:1180875.1180908,woodruff2007sabbath, 10.1145/2675133.2675282}. Offering a rich introduction to this category, Genevieve Bell's ethnographic essay \textit{No more SMS from Jesus: ubicomp, religion and technospiritual practices} critically interrogates the ways cultural and ritual practices shape people's relationships to technology in urban Asia based on research with 100 different Asian households~\cite{bell2006no}. Her work shows the increasing ways in which people are using technology to support their religious practices, emphasizing the ``need to design ubiquitous computing not just for a secular life, but also for spiritual life, and we need to design it now!''.  Finding that the two are already intertwined, she argues for the need to cater to spiritual expression online. She also suggests that current ubicomp development may inadvertently "preclude important spiritual practices and religious beliefs". Taking the cue from Bell's call, our inquiry unpacks the way Quran-related Islamic rituals are supported by technology, like Twitter, thereby, enriching this line of work with the case from the Arab Muslim context.

These studies, while limited in number within CSCW, put forward unresolved questions, tensions, and bi-directional effects regarding the role of technology in mediating rituals, especially with respect to their appropriateness in acts considered sacred. Susan Wyche~\cite{Wyche:2008:SDE:1358628.1358866,Wyche:2006:TSF:1180875.1180908} explores the question \textit{``What does it mean to design technologies that support religious practice?''} through the development of an application called ``Sun Dial'' that supports the Muslim daily prayer ritual by reminding users of prayer times. While convenient, this form of mediation still puts forward questions regarding ``the appropriateness of technology use in supporting religious rituals'': what is feasible, where are the boundaries, and when might these boundaries be crossed? To this point, a study by Rashidujjaman et al.\cite{10.1145/3025453.3025960} reflects on the techno-spiritual tensions experienced at the level of religious institutions (in this case, mosques in Bangladesh). In their work, the authors managed to help increase the mosque's donations through an SMS campaign to its constituencies. The study reveals tensions and nuance as age, technology type, and goal of mediation play a big role in determining technology's appropriateness. For example, while an older imam had reservations on the use of TV or cell phones, the younger one had fewer concerns regarding the use of Facebook to help reach the community.

Previous work also finds that automation\footnote{We define automation as computer programs developed to perform low-level functions like reminders or scheduling events at fixed times.} can play key roles in religious life. Woodruff's~\cite{woodruff2007sabbath} qualitative inquiry on how automation supports the Orthodox Jewish Sabbath ritual reflects on the benefits of technology in supporting an event that is centered in ``respite and renewal'' by performing mundane activities and offering a focused spiritual experience. Similarly, our study engages with forms of automation to support Islamic related rituals that are centered in reflection and contemplation.

 Other studies within CSCW and HCI show the significant ways Islam shapes the sociocultural practices and norms of Muslim people's online lives, activities, and practices and draw a holistic image for how technology used in everyday life (such as social media) could be appropriated to support a few Islamic rituals. A study by Ibtasam et al. \cite{10.1145/3359148} reveals the diverse and complex roles of different family members in introducing, teaching, and buying technology versus restricting and monitoring it in Pakistan. Individuals understood their roles in this regard as a function of their interpretation of Islamic beliefs (e.g., male guardianship). Alnasser's \cite{10.1145/2908805.2909426} research on designing ICTs to support positive ageing experiences in Saudi Arabia illustrates the religious aspect of technology use by the older people, for example, using WhatsApp to exchange morning and evening prayers. These behaviors are not unique to Islam, Ames et al.  \cite{10.1145/2675133.2675282} look at Judeo-Christian religious frameworks to uncover deep-rooted connections between engineering practices and discourses and religion as a general means of surfacing a few ``quasi-religious'' behaviors, such as practices of worship and evangelism in influencing people to convert to a specific approach or process (e.g., Design Thinking approach or volunteering at hackathons). These findings illustrate the ways in which religion and its concepts are deep-seated within cultures outside ones faith community including engineering cultures.

\subsection{Religion and Social Media}
 The second strand of literature centers the different ways religious communities utilize new media forms like social media as platforms for expressing, discussing, and practicing their religious identities~\cite{alshehri2018beauty,10.1145/3359163,10.1007/978-3-030-34971-4_2, armfield2003relationship,kluver2007technological,nyland2007jesus,campbell2012digital,campbell2005making,doi:10.1080/15348423.2014.971566}. We reference this work to better situate our study at the intersection of both lines of research on religion and technology.

 Early research on digital religion suggested that the internet offered users a space "for creating a personalized assemblage of their spiritual beliefs...[it also] offers an important opportunity for users to make visible their people's religiosity, since modern media usage allows them to create unique, dynamic presentations of traditional religious beliefs and formations."~\cite{campbell2012digital}. Looking at social media platforms today, similar opportunities arise. A study by Davidson and
Farquhar \cite {doi:10.1080/15348423.2014.971566} in the US found a positive correlation between religiosity, conservativeness, and Facebook-specific anxiety amongst Christians. The authors suggest that this is due to the nature of Facebook as a space of diverse beliefs and opinions puts religion users in a place where ``they have to negotiate these competing worldviews''. Relevant to this study, Al Hariri et al.\cite{10.1007/978-3-030-34971-4_2} study provides a rich analysis for 1.3M tweets of Arab users on Twitter discussing atheism and Islamic religion. They identified three categories of Twitter users according to the type of shared content: atheistic, theistic, and tanweeri (i.e. religious renewal). The discussions amongst these different groups varied between human rights discussions and national challenges facing Arab societies while all referenced Islamic content (i.e. Quran and Hadith) to either support or undermine the others' argument. These affordances and qualities of social media, while implicit, highlight their important role in creating an online space for discussing religious topics that were hardly possible in offline spaces given that some Arab countries criminalize atheism.

 Moreover, Kimmons et al. ~\cite{kimmonsreligious} explores religious expression on Twitter among Latter-Day Saints and identifies that they Twitter accounts tended to represent real identities (through real photos and real names) as opposed to anonymous identities or bots due to a sense of religious accountability. While some prefer using real identities, others use bots and automation apps for religious sharing on Twitter. Albadi et al's.~\cite{Albadi2018hate,10.1145/3359163} study found that bots were responsible for about 11\% of all religious hateful tweets on Arabic Twitter ~\cite{10.1145/3359163}. Their study revealed the use of 3rd party Islamic applications to post Islamic supplications and Quranic verses on their behalf. Similarly, a study by Abokhodair et al. ~\cite{abokhodair2015dissecting}, reports on the use of bots to automate and amplify the sharing of Quranic verses in Arabic for peace and stability during the Syrian civil war.  While these studies revealed traces of Quran sharing in their work, our study centers its investigation on this phenomenon, establishing by that a deeper engagement and documentation of Islamic related rituals enactment vis-a-vis technology.

Images also play a vital role in religious online life. A study by Alshehri ~\cite{alshehri2018beauty} explores the sharing of bloody photos on Twitter and Instagram during the Shia religious ritual Tatbeer -- the death anniversary of the prophet Muhammad's grandson -- to communicate with the local and global audience the authenticity and beauty of this religious ritual.

 Encompassing both literature categories is the work of Heidi Campbell's \textit{Considering Spiritual Dimensions with Computer-mediated Communication Studies} who explores how the internet is treated as a\textit{sacramental space}, where ``digital religion occurs'' by religious users through religious expression forming by that their religious identities, beliefs, and practice while also socially shaping the technology\cite{campbell2010religion}. Like Bell, Campbell underscores the importance of a deeper understanding of religious users and the ways their beliefs manifest online for a holistic view of religious technology uses. Overall, our study illustrates and further expands Campbell and others work by providing a case study from the Arab Muslim world, where Arab Muslims utilized and appropriated Twitter as a platform to support prominent Islamic rituals (e.g. reciting Quran and Charity) resulting in users engaging in religious expression and discussion via tweeting Quran verses. In addition, we provide nuanced meaning to a religious ritual and reflect on the hidden 'beauty' of sharing religious content that would otherwise be misunderstood or taken out of context.

\section{Quantitative Data Collection}
In the following section, we describe the engineering work performed for the data collection process of tweets consisting of Quran verses; we then discuss the verses categorization process applied to tweets for our analysis.

\subsection{Collecting Quranic Tweets Dataset}
The Twitter streaming API was used to stream tweets that contain key Arabic phrases typically associated with Quran verses; these phrases were particularly appropriate for the purpose of sampling Quran verses because they are typically used just before or after reciting a Quran verse. Table~\ref{tb:qurankeyphrases} shows the list of the seven different key phrases our team used to stream relevant tweets.  Given that sometimes Quran verses are mentioned without referring to these recitation phrases, our method nevertheless lent the data collection high precision for tweets with containing Quran verses. In addition, they do not introduce any bias to the topics of the collected Quran verses, since they are widely used irrespective of nature of a verse. A collection of 5.8 million tweets (not counting retweets) were collected over two years between January 1 2016 and December 31 2017. 

\begin{center}
\begin{table*}
{\footnotesize
\hfill{}
\begin{tabular}{p{4cm} p{9cm}} 
\textbf{Arabic key phrase} & \textbf{Meaning/Usage} \\
\hline
\<بسم الله الرحمن الرحيم> & means ``In the name of God''; used sometimes before stating Quran verse/s \\\hline
\<صدق الله العظيم> & used sometimes after stating Quran verse/s \\\hline
\<قوله تعالى> , \<قال تعالى> , \<قال المولى> , \<قال عز وجل> , \<قال في كتابه> & five different phrases that refer to the meaning ``God says'', which are typically used before stating Quran verse/s \\
\hline
\end{tabular}}
\hfill{}
\caption{Arabic key phrases used to stream tweets that potentially have Quran verse/s}
\label{tb:qurankeyphrases}
\end{table*}
\end{center}

Pre-processing for Arabic text was applied for effective string matching when spotting the Quran verses within the tweets. Basic pre-processing steps following~\cite{darwish2012language,darwish2014arabic} were applied, such as removing diacritics and kashidas that are optional in Arabic text, and  applying letters normalization by converting \{\<أ, إ, آ>\} to "\<ا>",  \{\<ؤ, ئ>\} to "\<ء>", "\<ة>" to "\<ه>", and "\<ى>" to "\<ي>" \cite{darwish2014arabic}. In addition, user name mentions (@user) and hash-tags were filtered out.  No stemming or stop words removal were applied, since we are interested in exact match of verses not text that uses Quranic terms but is not a verse.

\subsection{Presence of Quranic Content Validation}

\begin{algorithm}[t]
\scriptsize
\label{alg:validation}
Let \textit{T} denoted to a tweet which consist of a sequence of S sentences, \textit{S1},..., \textit{Sn}. We used a list of punctuations and newline to extract tweet sentence (\textit{S}).\;
 \KwData{Tweet sentences (\textit{S})}
 \KwResult{\textit{L} which is a list of Quran verses and it's categories which found in \textit{T}}
\While{not at end of \textit{S}}{
 read current\;
 \If{length(\textit{S})>2}{
 \If{\textit{S} in Quran\_Verses\_List}{
 check if \textit{S} is represented full verse or part of verses\;
 \eIf{\textit{S} represented the full verse}{
 get verse category from Quran\-by\-Subject dataset\;
 save verse and its category in \textit{L};
 }{
 get all verses which contain \textit{S} and its categories from Quran\-by\-Subject dataset\;
 save verses and its categories in \textit{L};
 }
  }
  }
 }
 \caption{Extraction of Quran Verses from tweets}
\end{algorithm}

The keywords used for collecting tweets are most likely used when a Quran verse or more exist. However, sometimes, they are used in another context. For example, the phrase ``In the name of God'' could be used at the beginning of statements or answers. In this step, we validate the content of the 5.8 million collected tweets for the presence of Quran verse or even part of a verse. We obtained a text version of the Quran in Arabic that is divided into chapters and verses. We applied the same text pre-processing steps applied to tweets.

A tweet text is split into sentences over punctuation and newlines; each sentence is then compared to the collected list of Quran verses, and the list of verses in a given tweet is retrieved. Since some of the Quran verses are long, where they are formed of multiple sentences and sometimes reach a full page, it is highly common to find only fragment of the verse shared on Twitter rather than a complete verse. Thus, we allowed spotting verse fragments within tweets. Our algorithm allows matching sentences of at least three words to be scanned over verses to detect any full or partial match to a verse. While some verses in Quran are only one or two word, we limited the shortest allowed sentences in tweets to three words to avoid matching popular phrases or stop words. In Arabic, three words phrase is unlikely to contain meaningless popular phrases due to the high inflected nature of the language, where pronouns and adjectives are connected to nouns and verbs~\cite{darwish2014arabic}. Algorithm 1 shows the steps for extracting Quran verse/s chunks from tweets and detecting its/their numbers.

After this step, the number of tweets that are validated to contain at least a fragment of Quran verse were 2.63 million tweets, containing a set of 3.4 million Quran verses (or fragments). We call this dataset \textit{Quran\_tweets}, and we use it for our analysis.

 To verify the accuracy of our algorithm, we selected a set of 100 random tweets that were identified to contain full verses and another 100 random tweets that contain verse fragments. One of the authors manually inspected them and found that all the 200 tweets contain verse/s (fragment/s) of the Quran. This shows that our algorithm achieves very high precision in detecting Quran verses and fragments in tweets.

\subsection{Quran Verses Categorization}

 Quran covers many topics from Resurrection and the Judgment, Paradise and Hellfire, to the early communities of faith and prophets. It also communicates Islamic laws (Sharia law) for personal affairs, society, and the state, including rules of marriage, family maintenance, inheritance, financial transactions, retribution, punishments for crimes, and many other topics \cite{nasr_2017}. For deciding the topic covered by a Quran verse, we use the categorisation presented by \textit{Quran-by-Subject}\footnote{ http://www.quranbysubject.com} website, which is the most popular Islamic website that categorizes the Quran verses into a set of thirteen categories according to topics\footnote{The website receives over 250K monthly visits according to SimilatWeb.com, and is ranked 301 worldwide in its category. Its application on Google play store has over 100K installs.}.  The website states that this categorisation to Quran was performed manually by Islamic scholars. We further validate this source by consulting a scholar in Quranic science about the website who confirmed its credibility.

These 13 categories are divided into a set of 171 subcategories. 4,912 of the Quran verses are manually assigned by experts to one or more of those categories, while 1,324 are not categorized into any. Table~\ref{tb:mainCats} shows the list of the main categories with the description of topics covered by each category and the number/percentage of verses in each. We added a fourteenth category "General" for those Quran verses that are not covered by any of the other categories. The sum of verses in Table~\ref{tb:mainCats} is larger than the number of Quran verses, since one verse can match multiple categories.
As shown in table, verses discussing the hereafter and stories of prophets constitute more than 50\% of the Quran verses. The remaining verses discusses topics such as disbelievers, Sharia law, Jihad, Worship, and others.

This categorization scheme was then used to label our tweets collection, where each tweet is labeled by all the categories of the verses it contains.

\begin{center}
\begin{table*}
{\footnotesize
\hfill{}
\begin{tabular}{l r r p{9cm}} 
\textbf{Categories} & \textbf{count} & \textbf{\%} & \textbf{Description}\\
\hline
Hereafter \& Unseens & 1,701 & 27.3 & Description of unseens, such as life in the hereafter, Paradise, Hellfire, Angels, creation of Adam, and devil\\
Stories of Prophets & 1,581 & 25.4 & Stories of the old prophets and messengers before Muhammad, such as Adam, Noah, Abraham, Lot, Moses, Jesus, and Mary\\
Disbelievers & 684 & 11 & Stories of old nations that disbelieved their prophets, and messages to disbelievers\\
Sharia Law & 487 & 7.8 &  Explicit statements for Muslim legislation in personal, social, economic, and political matters. Includes marriage, what is licit (halal) and illicit (haram), and ethical directives\\
Jihad & 397 & 6.4 & Jihad in Islam and its laws, and stories on battles during the life of Muhammad\\
Universe \& Creation & 388 & 6.2 & Examples of the creation of God, including the universe, stars, human being, animals, insects, and nature\\
Worship & 337 & 5.4 & Instruction of how to worship God, such as prayers, fasting, pilgrimage, repentance, and others\\
Belief \& Believers & 331 & 5.3 & Explanation to the belief in Islam, and description of believers and their manners\\
About Quran & 330 & 5.3 & About the Quran and its message\\
Muhammad & 326 & 5.2 & About Muhammad\\
God & 322 & 5.2 & Description of Allah (God) and his power and his graces to creatures\\
Sins & 98 & 1.6 & Examples of those who spread injustice and sins and explanation to their punishments in the hereafter\\
Human Being & 71 & 1.1 & The creation and nature of human being\\
General & 1,324 & 21.2 & Verses that are not included in any of the other categories\\
\hline
\end{tabular}}
\hfill{}
\caption{Fourteen categories that Quran verses cover, with the number and percentage of verses in each category. One verse of Quran can match one or more categories.}
\label{tb:mainCats}
\end{table*}
\end{center}

\section{Experimental Setup and Quantitative Analysis}
In this section we discuss the methodology of the quantitative analysis applied to the collected data and the experimental setup performed to further classify the tweets for an in-depth analysis.

\subsection{Detection of Automated Content}
When analyzing the tweets in our dataset, we found a trend of Quran tweets being tagged with the hashtag "\#\<دعاء> <du'a>" which translates to ``\#prayer'. We suspected that this trend might indicate bot activity, given that it was very consistent and frequent, which is also noted by ~\cite{10.1145/3359163}.  After further investigation of these tweets metadata, we found that these tweets were automatically generated through an app. We managed to identify 10 different applications \footnote{To name a few of these sources e.g., \url{https://du3a.org/}, \url{https://zad-muslim.com/}, and \url{http://alathkar.org/}} that users commonly use to automatically tweet Quran verses on their behalf. Users of these apps can set the frequency of tweeting, for example, they can set it for 1 verse a day or choose a certain Muslim event like Ramadan and Friday.  In this case, an account is considered a hybrid, where users can still post their own content in addition to the app content.

For a more precise analysis of human tweets, we separated all tweets in our collection generated by these applications. A set of 235K tweets that were posted by 24K users were generated by these applications. We call this set of tweets \textit{app\_tweets}, and since this is a significant number of tweets, we address this online behavior amongst Arab Muslims in the discussion section. The remaining set of 2.4 million tweets is named \textit{human\_tweets}, which we believe to be manually posted by users. Full statistics about these two sets are shown in Table~\ref{tb:qurancollection}.

\subsection{Influential Accounts Labeling}
Our dataset of 2.4 million \textit{human\_tweets} are posted by 696K different accounts. Out of these accounts we selected the most influential 500 accounts whose Quran tweets were retweeted the most in our collection regardless of the number of tweets they generated. These 500 accounts posted a set of 97K tweets containing Quran that have been retweeted 3.4 million times, which is more than one third of the total volume of retweets in our collection.

To answer our question \textit{who tweets Quranic verses?} and understand the different patterns in shared Quran subjects between the different types of accounts, we created a set of grounded categories in several rounds. In the first round, we selected a little over 100 tweets at random from the 500 accounts to generate the initial codes. One coder (a native Arabic speaker), read through the tweets and proposed a draft coding manual. After that the categories and tweet exemplars were reviewed and discussed by the authors. In a second round, we took a random selection of tweets to test and elaborate the categories from the first round leading to couple of revisions of the initial set of codes. Finally, 50 accounts (from the influential accounts) were annotated by the three authors (all are native Arabic speakers) for measuring inter-annotator agreement resulting in Cohen's Kappa of 0.76 and 0.73 for the main and secondary categories receptively, which indicates a reasonably high agreement among annotators.

 At the outset, two overarching coding categories were identified: 1. Main categories are based on whether the account represents a page or a person, 2. Secondary categories are based on the type of the shared content (only religious content, or content that is generally non-religious). The goal of this secondary level of coding, is to have a high level understanding of the types of accounts that only share Quran, dua, Hadith. The following guidelines were provided for annotators for the annotation process: 

\textbf{Main categories}
Since we isolated \textit{app\_tweets} from this dataset, we focused the analysis on two main labels: Person and Page.
\begin{itemize}
\item \textit{Personal}. An account is labeled as a person (as opposed to a page or a bot) when the following criteria are met:  1. account screen name is a real person name and not a nickname (e.g., "\textit{Dr. XYZ lovers}"). 2. profile picture resembles a real personal photo\footnote{this is more difficult to achieve in this dataset due to the large number of accounts with no profile picture for cultural reasons ~\cite{abokhodair2017photo}}. 3. profile bio and features look to refer to an individual, such as clear social status or occupation (e.g. husband, wife, doctor, teacher ...etc.). To ensure a high validity of our labels, we followed recommendations from \cite{Albadi2018hate} on labeling Arabic human versus bots accounts.

\item \textit{Page}. An account is labeled as a page when one of the following criteria is met: 
1. all the above mentioned conditions are not met. 2. any indication to the account being a page representing a business, an interest-group, a government agency or a ministry, or celebrities from the account name or description\footnote{We also checked for potential bot accounts, but none of the labeled accounts had a strong evidence of being a bot. The only noticed bot-like activity was the presence of automatically generated tweets using the applications mentioned earlier, but it was clear that these accounts still post other manually generated tweets.}
\end{itemize}

\textbf{Secondary categories}
Additionally, we developed the two main labels based on profile purpose and tweet content. These codes are independent from the primary codes:
\begin{itemize}
\item \textit{Religious-Content-Exclusive (RCE)}. An account is labeled RCE when the following criteria is met: 1. profile biography indicates that the account serves only religious purposes, like a sharing Quran verses, dua, and prayer reminders, or a profile for a Muslim preacher/scholar; and 
2) tweet content is mostly religious, including religious advice, Hadith, and Quran verses\footnote{For the purpose of this study, we focused our analysis on identifiable religious content like Quran, Hadith, and Dua. Content that carried religious attitude, tones, endorsement, or debate is out of scope for this paper}. 
\item \textit{General}.  An account is labeled general when following criteria is met: if the criteria above are not met, for example the account posts about other topics, such as, sports, politics, business, etc. with scattered religious content in the timeline. In this case it is coded as general.
\end{itemize}

The 500 accounts were annotated as 72 personal-RCE, 114 page-RCE, 194 personal-general, 120 page-general. Table~\ref{tb:qurancollection} shows full statistics on the number of tweets and verses posted by these accounts. An in-depth analysis is performed to compare the Quran content shared by these accounts. Furthermore, the manual annotation reveled some intriguing trends, such as Twitter accounts created for deceased people that seemed to be managed through third party applications to tweet on behalf of the owner. These observations motivated us to run a follow up qualitative study with Twitter users to better understand these trends and the motivations behind these online behaviors.

\subsection{Quantitative Analysis}
In the first part of the analysis, we measure the engagement with tweets containing Quran verse/s to see the likelihood of getting these tweets retweeted. In this case, we calculate the distribution based on the total of retweets rather than tweets, since that type of engagement represents the volume of interest in a topic more than the tweets count~\cite{gao2014effective}. Our hypothesis states that if a tweet with a given verse receives 1000 retweets, it can be considered more popular than 100 tweets with the same verse but never been retweeted. While retweet count is more reasonable to be used for calculating distribution, there is a concern that the number of retweets might be affected by the influence of the account tweeting the verse rather than the verse content itself. To inspect this concern, we calculated the correlation between the full number of retweets of Quran tweets posted by a given account and the number of followers of this account. The resulting Poisson correlation was 0.103, which indicates independence between the sum of retweets of a specific Quran verse and the account's influence. 

For the second analysis, we analyze the distribution of topics stated in these tweets to understand whether the distribution of shared topics align with the distribution of topics in the Quran, or people focus on specific topics over others when sharing Quran on social media. As mentioned earlier, many verses of Quran are labeled with more than one topic. In our analysis, the distribution of categories would be calculated based on the percentage of verses that match the categories. The percentage of category $C_i$ in the volume of retweets is calculated as follows:

\begin{equation}
Percentage(C_i)=\frac{count(verses_i)}{count(verses)}
\end{equation}

Where $count(verses_i)$ is the count of verses in all tweets and retweets that matches category $i$, and $count(verses_i)$ is the total number of verses in all tweets and retweets. For example, $Percentage(C_{worship})$ = 18\% means that 18\% of the verses in the full volume of tweets (including retweets) in our collection matches this category; this does not contradict that some of these verses still match other categories.

Next, we report on the results of the quantitative analysis from both \textit{human\_tweets} and \textit{App\_tweets}, including the most retweeted Quran verses (complete and incomplete) and explain the main themes covered by them.
\begin{center}
\begin{table*}
{\small
\hfill{}
\begin{tabular}{llrrrrr} 
\hline
\textbf{Dataset} & \textbf{posted by} & \textbf{\# accounts} & \textbf{\# tweets} & \textbf{\# verses} & \textbf{tweets volume} & \textbf{verses volume}\\
\hline
\textit{human\_tweets} & People/pages & 696,342 & 2,399,588 & 3,174,101 & 9,076,861 & 10,823,860 \\
\textit{App\_tweets} & Apps & 23,968 & 23,4622 & 23,4775 & 24,6355 & 24,8176\\

\hline
& Personal-RCE & 72 & 16,799 & 23,389 & 566,744 & 681,305\\
labeled & Page-RCE & 114 & 56,758 & 62,049 & 1,055,055 & 1,353,921 \\
accounts & Personal-general & 194 & 20,505 & 21,193 & 1,008,579 & 1,149,441\\
& Page-general & 120 & 3,341 & 5,313 & 781,399 & 879,803\\
\hline
\end{tabular}}
\hfill{}
\caption{Full statistics on our \textit{Quran\_tweets} collection and the labeled influential accounts. Tweets volume indicates the total number of tweets and retweets. Verses volume is the total number of verses in the full tweets volume}
\label{tb:qurancollection}
\end{table*}
\end{center}

\section{Quantitative Analysis Results}
\label{sec:results}

\subsection{Quran Tweets Statistics}
Table~\ref{tb:qurancollection} reflects the full statistics of our collection for both \textit{human\_tweets} and \textit{App\_tweets} datasets. The average number of verses per tweets in the \textit{human\_tweets} set is 1.36, while it is almost 1 for application-posted tweets. This shows that \textit{App\_tweets} consistently have only one verse/tweet (or part of verse), while human tweets tend to share more than one verse in the manually shared tweets. This is expected since these apps are programmed to tweet one verse at a time, whereas human posted tweets have a different pattern depending on the motivation. 
The full tweet volume of \textit{human\_tweets} when counting the retweets is 9M, which indicates an average of 2.78 retweets per tweet. While for \textit{App\_tweets} the volume of tweets is almost the same as number of tweet, showing  that users are less likely to retweet automatically generated Quran verses (through bots or applications).

\begin{figure}
\center
\includegraphics[width=0.5\textwidth]{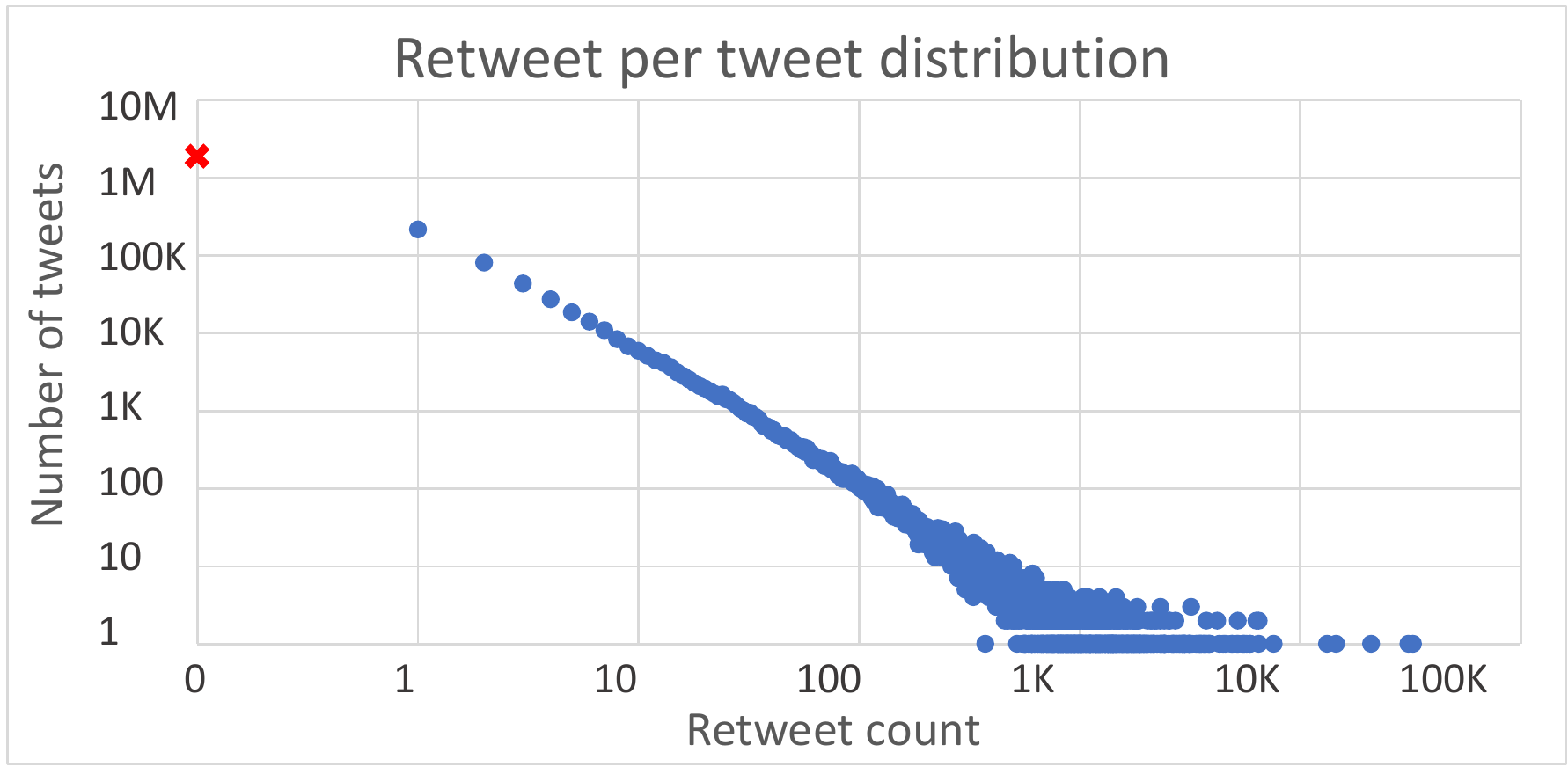}
\caption{The number of retweets received for tweets in \textit{human\_tweets} dataset. Graph is plotted at a log-log scale}
\label{fg:retweet_dist}
\end{figure}

Moreover, the average retweet per \textit{human\_tweets} is 2.78; however, not all tweets are retweeted equally. Figure~\ref{fg:retweet_dist} shows the retweets distribution for the \textit{human\_tweets}. Only 22\% of the tweets get retweeted, which is still significantly higher when compared to 6\% for \textit{App\_tweets}. As shown in Figure 1, the retweet pattern follows a Zipf distribution, which is a popular phenomenon in many aspects of life, including posts share in social media as reported in ~\cite{giatsoglou2015retweeting}.

\subsection{Categories Distribution of Shared Quran Verses}
Using the set of categories listed in Table~\ref{tb:mainCats}, we calculate the distribution of Quran topics that are shared on Twitter. First, we provide the distribution of topics in both the \textit{App\_tweets} and \textit{human\_tweets} data sets and compare it to the distribution of the same categories in the Quran itself. Secondly, we calculate the differences in topic distribution for the most influential accounts by type. As mentioned earlier, the goal of these two analyses is to better understand the behaviors around sharing Quran verses on Twitter and whether certain Quran topics are retweeted more than others.   

\subsubsection{Topic distribution in Quran verses vs. Twitter}

\begin{figure}
\center
\includegraphics[width=\textwidth]{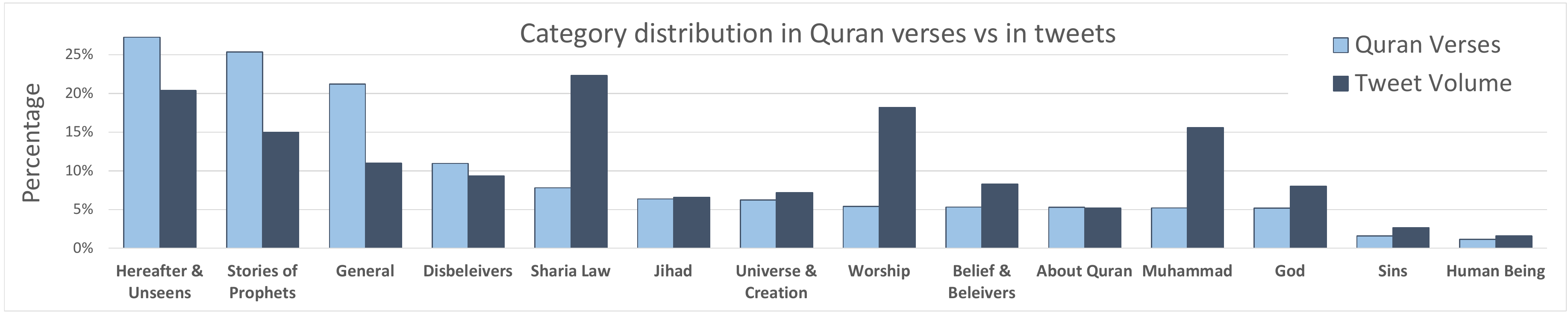}
\caption{Categories distribution of topics discussed by Quran verses in Quran itself vs in the full tweet volume of \textit{human\_tweets} dataset (9 million tweets/retweets)}
\label{fg:DistQuranTweet}
\end{figure}

By inspecting the topics of verses in \textit{App\_tweets}, we found that they mainly cover one category that is \textit{Muhammad} (94\% of the tweets are within this category). When looking closely at the data, we discovered that one verse is repeatedly shared by these apps, i.e. (33:53) in Table 4. Likewise, the same verse is one of the most shared verses in the human\_tweets collection. This shows that these applications are very basic, since they share similar content regularly, which might further explain our earlier finding that they are seldom engaged with through retweeting. 

Figure~\ref{fg:DistQuranTweet} plots the distribution of the 14 categories covered in the Quran in comparison to how they get shared on Twitter from the \textit{human\_tweets} dataset. Figure~\ref{fg:DistQuranTweet}, demonstrates clear differences regarding the emphases of sharing certain categories on twitter in comparison to the emphases in Quran. One key observation regarding the category distribution in the Quran shows that the two most covered topics in Quran are \textit{Hereafter and Unseens} and \textit{Stories of Prophets}. Differing from the 3 topics that get mostly covered on Twitter, those are \textit{Sharia Law}, \textit{Worship}, and \textit{Muhammad}.

\subsubsection{Page vs. Human Accounts}

\begin{figure}
\center

\includegraphics[width=\textwidth]{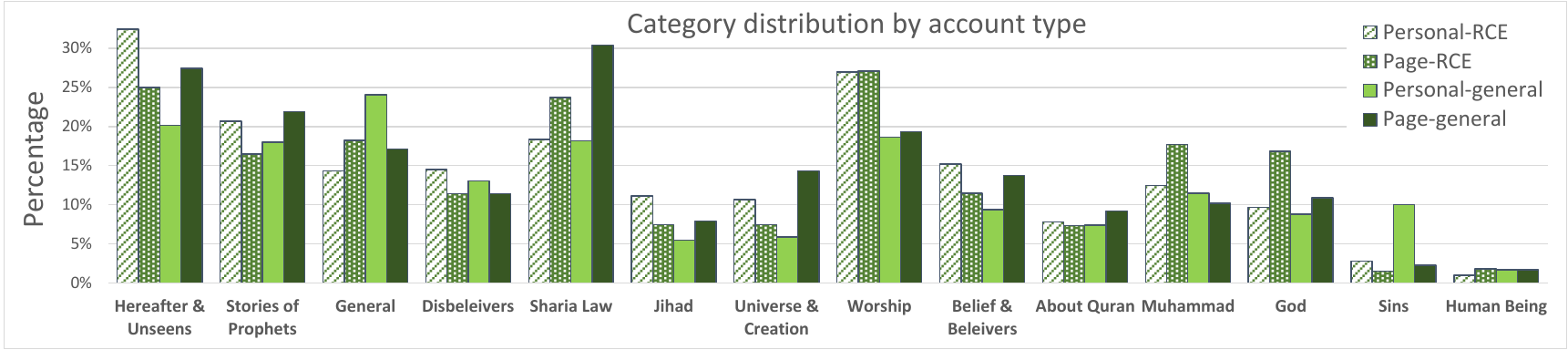}
\caption{Categories distribution of Quran verses shared in the full tweet volume (3.4 million tweets/retweets) posted by the most influential 500 accounts, compared by account type}
\label{fg:DistAccounts}
\end{figure}

Taking our analysis one level deeper, Figure~\ref{fg:DistAccounts} displays topic distribution in the tweets of the most influential accounts according to their type. Overall, the trend of topics aligns with the trend of the full tweets volume of \textit{human\_tweets} in Figure~\ref{fg:DistQuranTweet} with some differences between account types. One Key observation shows that personal RCE accounts and Page RCE tweet mostly about \textit{Hereafter \& Unseens} and \textit{Worship}.  The only clear difference is that Page RCE mostly tweets verses about \textit{Sharia Law}. This is due to the fact that most Page RCE account are for known imams and religious institutes that have the responsibility of educating Muslims on the everyday personal and social affairs, and share knowledge regarding Islamic laws and instructions that include everyday life matters, like marriage, inheritance, ethical directives, prayers, etc. In this case, Twitter is seen as a place where Muslims and Muslim scholars can reach a wide number of followers to remind and educate each other on rituals and matters through using Quran verses.

In addition, we observe that local business pages that are not dedicated to religious content (Page-general) share Quran verses, mostly in the category of \textit{Sharia Law} and \textit{Hereafter}, however, periodically. We found that these tweets receive large engagement by followers through retweets and likes.  We hypothesis that the use of Quran verses in this case is mostly to gain religious favor with the audience and to reflect high engagement numbers (e.g., click bait purposes), which could mostly be generated by bots. However, a deeper analysis into these retweets is out of scope for this paper and can be addressed in future work.  

Undoubtedly, these different quantitative findings reveal that Quran verses are commonly shared by different types of Twitter accounts with some differences in the sharing patterns, which depends on the account main interest or business.

\begin{center}
\begin{table}
\footnotesize
\begin{tabular}{l p{12.5cm}} 
\hline
\textbf{Verse} & \textbf{English Translation}~\cite{nasr_2017}\\
\hline
(48:1) & Truly We have granted thee [Prophet Muhammad] a manifest victory.\\
(4:110) & Whosoever does evil or wrongs himself, and then seeks forgiveness of God, he will find God Forgiving, Merciful.\\
(33:56) & Truly God and His angels invoke blessings upon the prophet. O you who believe! Invoke blessings upon him [Prophet Muhammad], and greetings of peace!\\
(68:4) & And truly thou art of an exalted character.\\
(113:1) & Say, "I seek refuge in the Lord of the daybreak".\\
(113:5) & and from the evil of the envier when he envies.\\
(113:2) & From the evil of what He has created,\\
(113:3) & From the evil of darkness when it enshrouds,\\
(112:1) & Say, "He, God, is One,\\
(112:3) & He begets not; nor was He begotten.\\
\hline
\end{tabular}
\caption{Top 10 shared Quran complete verses in \textit{human\_tweets} collection as translated by ~\cite{nasr_2017}. Original Arabic verses are provided in the appendix in Table~\ref{tb:topCompletear}}
\label{tb:topComplete}

\begin{tabular}{l p{12.5cm}} 
\hline
\textbf{Verse} & \textbf{English Translation} ~\cite{nasr_2017}\\
\hline
(65:1) & Thou knowest not: perhaps God will bring something new to pass thereafter.\\
(19:64) & and thy Lord is not forgetful--\\
(8:33) & And God will not punish them while they seek forgiveness.\\

(26:227) & And those who do wrong shall know to what homecoming they will return.\\
(40:21) & yet God seized them for their sins, and they had none to shield them from God.\\
(7:100) & if We willed, We could smite them for their sins and set a seal upon their hearts such that they would not hear?\\
(19:21) & He said, "Thus shall it be. Thy Lord says, 'it is easy for Me.'"\\
(5:32) & and whosoever save the life of one, it is as though he saved the life of mankind altogether.\\
(7:156) & though My mercy encompasses all things. \\
(14:7) & If you are grateful, I (God) will surely increase you.\\
\hline
\end{tabular}
\caption{Top 10 shared Quran verse fragments in \textit{human\_tweets} collection. Original Arabic verses are provided in the appendix in Table~\ref{tb:topIncompletear}}
\label{tb:topIncomplete}
\end{table}
\end{center}

\subsection{Top Shared Verses}
We conducted an additional analysis to answer the question \textit{what are the most shared Quran verses and what are they about?} In doing so, we extracted the top 10 shared Quran verses in our dataset, whether the tweets contained complete or fragment Quran verses. It is worth noting that Quran verses can be unspecific and metaphorical. Therefore, it is on the reader to analyze both the context and the verse to determine if it is metaphorical or literal\cite{nasr_2017}.  To aid in this process, scholars have created rules for deciding which phrases are metaphorical and which ones are not that are written in exegesis books (Tafsir in Arabic). To share examples of the verse analyzing process in context, we highlight few verses from the list and discuss their literal and metaphorical meaning from \cite{nasr_2017}\footnote{"The Study Quran" book~\cite{nasr_2017} is a highly referenced English translation of the Quran and its tafsir produced by a distinguished team of Islamic studies scholars. It has more than 420 4+ ratings online and more than 170 citations} to hypothesis the sharing motivation and behavior.

\subsubsection{Top 10 shared complete verses}
Table~\ref{tb:topComplete} lists the 10 most shared complete verses (translated to English) by~\cite{nasr_2017}.  The most shared verse across \textit{human\_tweets} is the opening verse from the Victory chapter (48:1) that literally describes Muslims victory in the conquest of Makkah~\cite{nasr_2017}. However,the meaning of the verse is not only concerned with victory in war, it is also a metaphor to describe the opening of the heart and the mind to "the unveiling of the secrets of the Divine Essence"\cite{nasr_2017} with optimism for a future success \textit{"a manifest victory"} (48:1). This verse falls within the \textit{Muhammad} category, Table~\ref{tb:mainCats}, since it was addressing Muhammad in the speech. In reading a few tweets referencing this verse, we hypothesis that the high volume of sharing this verse is mainly for its metaphorical meaning that incites hope and optimism, especially during tough times like war and injustice. 

Second in the list, is a verse on  God's forbearing generosity and expansive mercy (4:110) from An-Nisaa chapter that falls within the \textit{Worship} category. Seeking forgiveness is an act of worship in Muslim faith and the verse verbatim is commonly used amongst Muslims as a reminder of the limitless mercy and forgiveness for those feeling remorse. This can be the reason why people highly share this verse on Twitter.

The third and fourth ranked verses --the former from the 33rd chapter named The Parties, and the latter from the 68th chapter named The Pen-- and both verses fall in the Muhammad category because they provide a description of his elegant morals (68:4) and a reminder for sending blessing upon him (33:56). These verses serve as a reminder for Muslims to ``\textit{invoke blessings} by supplicating for the Prophet''~\cite{nasr_2017} and to follow his morals in their everyday life.

The last six verses in the list are from the two shortest and last chapters in Quran, the Daybreak (113) and Sincerity (112). Muslims recite these chapters as part of a daily morning ritual for protection and blessings ~\cite{nasr_2017}. The Daybreak chapter is formed of five short verses and is known as one of the "Two Protectors" that star with the verse "say: I seek refuge in the Lord of...". The Sincerity chapter consists of only four verses and is one of the few chapter that provides Muslims a description of God. It is of high importance "in Muslim devotional life and is often recited in both canonical and supererogatory prayers as well as in dua" ~\cite{nasr_2017} because of its focus on Monotheism, the first pillar of Islam.  We anticipate that these verses appear together in the same tweet rather than separate verses\footnote{Verse (112:2) is missing, since it contains only two words and got filtered out by our algorithm that filters verses less than three words} and are used from the same purpose on Twitter, mainly as a reminder for starting one day with the blessings of these chapters.

\subsubsection{Top shared verse fragments}
Table~\ref{tb:topIncomplete} lists the most popular 10 shared verse fragments in our dataset (translated to English) by~\cite{nasr_2017}.  As mentioned earlier, the significance of verse fragments is that they are mostly shared for their metaphorical meaning to serve a purpose that is not similarly served when the full verse is shared as demonstrated in the top shared examples.

The top ranked verse fragment (65:1) focuses on dealing with uncertainty. This is the last sentence in a much longer verse on the topic of divorce that falls within the Sharia law topics. This verse has been appropriated in daily language to remind people to stay positive and hopeful and that no pain lasts forever\cite{nasr_2017}. So, while the whole verse topic is on the laws and ethics of Islamic divorce, the use of this fragment on Twitter is mostly for its metaphorical meaning that is an advice to have continues patience in dealing with uncertainty. Similarly the seventh top ranked fragment (19:21) is part of a verse about the miraculous birth of Jesus where the phrase is the reply by an angel when Mary asked how she would be pregnant without having a human touching her~\cite{nasr_2017}. Again, this phrase is used by Muslims in hardship to remind them of God's ability in relieving hard situation.

The second most retweeted verse fragment is (19:64) which serves as a reminder to Muslims that God never forgets them. It is commonly used by Muslims when going through a situation of injustice to remind that the God is not forgetting about their hardship, which we suspect to be the same reason for being shared on Twitter. The same theme is covered in the fourth (26:227), fifth (40:21), and sixth (7:100) top ranked ones, where they all discuss the same topic, God's power and ability to punish those who oppress other and cause harm/pain to humankind.  We suspect that all these fragments are shared for the same purpose on Twitter, solidarity and support. The third  (8:33) and ninth (7:156) verses discuss the mercy of God, while the eighth (5:32) encourages Muslims to save lives and be helpful to mankind. The tenth (14:7) encourages Muslims to be grateful to God and to those around them highlighting the value of gratitude in everyday life~\cite{nasr_2017}.

\subsubsection{Top Verses by Topic}
Additionally, we analyzed the top shared verse in each of the 14 Quran categories and found that they follow the same theme in the top shared verses (complete and fragment). Within the \textit{Jihad} category, the top shared verse is  "\textit{Say, "Naught befalls us, save that which God has decreed for us. He is our Master, and in God let the believers trust.}(9:51)" The literal meaning of the verse is concerned with reminding Muslims that their fate and destiny is determined by God, thus no harm is upon them unless it is decreed \cite{nasr_2017}. Muslims usually reference this verse as a reminder to trust in fate and not fear any upcoming event that might be stressful.

This section highlighted the most shared Quran content on the Arabic Twittersphere by category and by account type. As we looked deeper into the meaning of the verses, we observe that most of these verses center around themes of mercy, helping others, forgiveness, solidarity, and the aim for a better future.

 In concluding our quantitative analysis, we answered our sub-research questions related to the how people share Quran on Twitter and what is being shared of it. As we continued this analysis we had many speculations regarding \textit{why do people share Quran verses on Twitter}, \textit{why do people retweet Quran tweets tweeted by a human and not retweet application-generated Quran tweet?}, \textit{What is the trend of creating Quran tweeting accounts for the deceased on Twitter?} Intrigued by all these questions and eager to provide a holistic view of this practice, in the following section, we discuss our qualitative study methods and findings. 

 \section{Qualitative Study}
 To answer our research questions and to better understand the role online expression play in religious life for Muslims from the Arab world, we designed and conducted a qualitative study that consisted of an online survey on Twitter followed with in-depth interviews.
\subsection{Survey study}
The first author, who is a native Arab Muslim, designed a survey\footnote{Translated survey instrument attached to appendix} instrument in Arabic with questions to better understand the behaviors and attitudes towards Quran sharing on Twitter. To cast a wide net of Arab Twitter respondents, we administered the survey through email and on Twitter via all the authors personal accounts. To broaden participation in the survey, we asked respondents to retweet a link to the survey through their accounts. The downside of this method was concerning reaching enough respondents with diverse backgrounds, opinions, and responses.  This was mitigated by asking couple of Arab public figures on Twitter to retweet the online questionnaire. There were no pre-set conditions on background, gender, or location to responding to the survey. The only two conditions were that the age of the respondent is above 18 years-old and that the respondent has a profile on Twitter.
. 
\subsubsection{Survey Instrument}
 The online survey consists of 13 questions, and took an average of eight minutes to complete. The survey instrument included a collection of closed-ended and open-ended questions, mostly inspired by the questions we had from the quantitative analysis findings. The survey instrument was piloted with five Arabic-speaking researchers in the field of social computing for questions formatting, clarity, and order. After a few iterations based on feedback, the questionnaire was finalized and organized into four parts. The first section, related to the subject's demographic information: age, gender, and location (country). The second section asked questions about Twitter use and frequency of use. The third section consisted of questions related to attitudes towards sharing Quran verses on Twitter, covering motivations, engagement and reaction, frequency, and modality of posting (manually compose or via apps to automatically generate them). Finally, we included a question to recruit respondents for a follow up semi-structured interview study.

 \subsection{Respondents}
We received 135 completed surveys after four weeks of administering the survey. We stopped collecting more responses when we deemed that data saturation was reached in the answers to the open-ended questions (i.e., most themes were repeating in the data). While we estimated that saturation was reached after 90 responses, we continued with a few additional responses to confirm that no new themes were emerging and ensure enough diversity within participants' demographics. Before analyzing the data, we ensured cleansing the results from any blanks and incomplete answers. After that, we conducted a thematic analysis to characterize respondents answers to the open-ended questions.

\begin{figure}
\center
\includegraphics[width=0.95\textwidth]{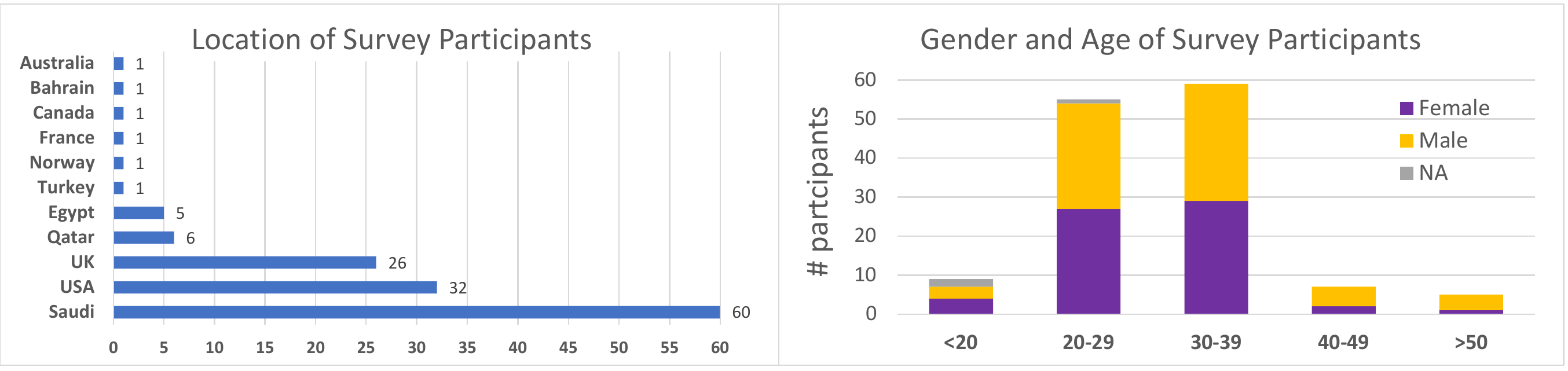}
\caption{Demographics statistics of survey participants}
\label{fg:survey}
\end{figure}

Figure~\ref{fg:survey} shows the demographic statistics of the survey participants. As shown, we had participants from 11 different countries, with 85\% of them coming from three countries: Saudi Arabia (60 participants), the United States (32), and the United Kingdom (26). Those living in English-speaking countries were Arab Twitter users, since the survey and the provided answers were all written in Arabic. In terms of gender and age, we had a close ratio of male to female participants (M=69, F=63), and 3 preferred not to record their gender. 86\% of participants were between 20-39 years old.

\subsection{Twitter Usage and Quran Sharing}
In terms of usage patterns, the majority of participants said they frequently used Twitter for reading others tweets (64\% daily usage and 15\% 3-4 times a week), while only 8\% used it 1-2 times a week, with 16 respondents rarely using it. On the other hand, the vast majority of respondents said they tweeted 1-2 times a week (54\%) and 15\% reported tweeting daily. 13\% tweeted 3-4 times a week, and 18\% said they rarely tweet. The majority of respondent (78\%) indicated encountering Quran verses on Twitter and 47\% (n=65) said they themselves tweet Quran verses, with a varying frequency of sharing. 28\% reported they tweeted Quran verses once a week, while 9\% shared it twice a week. The majority of those who share the Quran on Twitter (68\%) said they always manually compose their Quran verse tweets, while 17\% indicated hybrid sharing, with only 5 respondents indicated they only use automatic applications for tweeting Quran verses. 

 As the results reveal, about half of our respondent sample indicated they share Quran on Twitter while the vast majority have encountered Quran verses on Twitter confirming a common trend of sharing Quran on Twitter within our respondent sample. That said, a large number of respondents said they never shared Quran verses on Twitter (52\%).

63\% respondents answered the open-ended questions and 13 participants indicated interest in participating in a follow up interview, out of which, we managed to follow up with seven respondents through semi-structure interviews. 

 \subsection{Interviews}
 \subsubsection{Participants}
 The first author initially recruited study participants from the survey respondents who indicated interest in participating in a follow up interview (total of 5 accepted participation), who then recommended additional participants through their personal and professional network. The downside of this method was concerning reaching a large enough number of participants with diverse backgrounds (most interview participants were from Saudi Arabia). That been said, we acknowledge that the recruitment process is subject to self selection bias and thus our findings might not be representative of the whole population of Arab Muslims on Twitter. Our sample for this research is comprised of seven participants, three males and four females. In addition, most of the participants were between 25 and 35 years old (N=5), who identified as either higher-education students (Masters or PhD) or professionals with full-time jobs. Participants were offered compensation for their time, in the form of a small amount gift card, but most participants were happy to take part in the study voluntarily and rejected monetary compensation.
 
 \subsubsection{Interview Method}
 Living and experiencing daily interactions first hand as researchers from an Arab Muslim background gives us insight into the nuances and complexity of the religious and cultural practices of the region. The first author, an active researcher on the topic of Arab Muslim online identity \cite{abokhodair2017transnational, abokhodair2017photo,Abokhodair:2016:PSM:2901790.2901873,al2017against} engaged in conversational interviews, also known as \textit{ethnographically informed interviewing} with seven participants. This type of interviewing offered us flexibility in the flow of questions, and provided a space for our participants to feel free to share in-depth accounts of their experiences. When conducting this type of interview, it is also expected that "interview questions will change over time, and each new interview builds on those already done, expanding information that was picked up previously, moving in new directions, and seeking elucidations and elaborations from various participants"\cite{patton2005qualitative}. This approach to interviewing was especially valuable in addressing some of the observed trend from the survey with the participants.

Interviews were conducted through Microsoft Teams, in Arabic (a native language for the authors and participants); each lasted from 60 to 90 minutes on average. Code-switching did occur, but Arabic remained the language that was primarily spoken\footnote{Code-switching refers to the transitioning between languages in a single conversation among bilingual speakers}. During the interviews, we started with high-level questions such as "how often do
you check check Twitter?", "what types of accounts do you like following?", "how do you react when you encounter a Quran verse in your Twitter feed?", and "do you share Quran verses on Twitter?". From there, participants shared their thoughts and feelings regarding Quran sharing with little prompting; we were careful to listen, to allow them to express themselves freely, and to ask probing questions as they naturally came up in conversation. This led to rich and varied discussions that provide the foundation for our major findings.

\subsection{Qualitative Data Analysis}
Survey open-ended answers were cleaned from blanks and incomplete sentences, then included in a document for analysis. Separately, all seven interviews were transcribed in Arabic by the first author who then open coded the transcripts (following the protocol described in \cite{seidman2006interviewing}). Overall, our qualitative dataset consisted of the combined survey open ended answers and the interview transcripts, since the survey open-ended questions were a subset of the questions answered during the interviews. 
The generated codes were then carefully translated to English for reporting during the axial coding stage, memo writing, and clustering process. During the transcribing and open coding stages, we took into consideration the context by explicating the meaning in relationship to the context rather than a mere focus on the literal. This was especially the case when idioms and religious symbolism were used or terms that have a certain meaning in Arabic dialogue. Being "cultural insiders" is an advantage in this situation as it allows us to better explicate and explain the values as commonly practiced in Arab Muslim societies. After open coding, affinity diagraming was used to enact constant comparison between open codes to generate more thematic clusters. We generated themes based on our readings of the whole group of answers and triangulated results with each other throughout our analysis. In this work, we focus on themes most relevant to the perceptions and practice of Quran sharing on Twitter.

\section{Qualitative Findings}
The overarching theme we observed from our qualitative analysis was that Arab-Muslim Twitter users engage in Quran sharing practices to support them in practicing specific common rituals that were often enacted offline, however, found new expression online. We show evidence of how Twitter is conceptualized as a \textit{virtual sacramental space} that is reshaping what it means to be religious in society and as a result reshaping offline religiosity\cite{campbell2005considering}. Here we detail some of these rituals and discuss what they achieve for our participants and survey respondents. In doing so, we offer answers to "what role does online expression play in religious life for Muslims from the Arab world?" and unpack the affordances within social media technology to support the practice of certain Islamic rituals online. We present our themes into five major clusters: ceaseless charity, remembrance agents, expressions of solidarity, digital memorials, and Quran automation on Twitter.

\subsection{Ceaseless Charity -- Sadaqah Jariyah}
Whether alive or deceased, sharing Quran verses on Twitter to gain the rewards of those who read it and engage with it motivates people to share Quran verses on Twitter. In Islam this practice is referred to as ceaseless charity (\textit{sadaqah jariyah} in Arabic) and is concerned with the notion that any good deed a Muslim person starts during their lifetime that is of renewed benefit and ongoing use for other Muslims or humanity will continue to benefit them and augment their record of good deeds, even after their death --as long as the benefits continue to reach others ~\cite{kauthar_2017}. Most interview participants and 30\% survey respondents referenced the continuous accumulation of good deeds during and after their life as the motivation for sharing Quran verses on Twitter. This is illustrated in the answer of a male interview participant (M-3) regarding why they share Quran on their Twitter account, \textit{"If I feel the meaning of the verse and share it on my page, I hope that I'll gain the good deed of someone else encountering it on my page and giving them the same feeling"}. A survey respondent answered that sharing Quran verses on Twitter helps, "\textit{[to] bring me closer to the Creator [Allah] and a good word to leave behind me (after I die)}". Likewise (SR-112) "\textit{to be remembered positively when someone visits my account and for the reward of reminding people}". We note the extended interpretation of the word "charity" from the hadith to mean more than just the common notion of monetary support to include 'a good word'. For many Muslims, that 'good word' is translated to mean the words of the Quran, hadith, or dua. This example illustrates the new meanings associated with the evolution of rituals from offline to online spaces.  

Arab Muslims conceptualize social media platforms like Twitter as everlasting \cite{abokhodair2017transnational}, at least beyond their life time, therefore, users find them an ideal channel for practicing the ritual of ceaseless charity online, which is a very common offline ritual as represented by female interview participant (F-2), "When I started my Twitter account I wasn't sure what to tweet, I wanted to have content that is meaningful and educational, something to be remembered by and for people to benefit from. I started tweeting Quran verses to remind people of Quran and gain the rewards." Similarly, (SR-43) said, "\textit{To gain the deed for myself and others who read it. It is also considered a ceaseless charity}". Since the ritual of ceaseless charity is focused on educational and inspirational material, understanding it helps explain the topic trends of the most shared Quran verses. For example, looking at the high trend of sharing of sharia law or worship, can be motivated by this practice since both topics include teachings of Islamic rituals and way of life. Commenting on this, an interview participant (F-3), who had her account for more than 10 years, said "I'm very selective of the verses I tweet, I want them to have value for my audience, who is not all Muslim, and also be easily translated by Twitter, for those who don't read Arabic. So, I share verses on the beginning of holidays, such as Ramadan, Hajj, and Friday, for the simple reason of educating people on these holidays." F-3 is intentional in her sharing behavior to reach out not only her Arabic-reading audience, but also English speakers who might use the translation feature to the content of her tweet. In her intention to educate and practice ceaseless charity, the verse choices she made (i.e. holidays and Islamic events) is a topic that falls under both sharia law and worship categories. Providing an additional explanation to the spike in sharing these categories.

 \subsection{Reflection, contemplation and remembrance agents}
  Another Quran-related ritual is concerned with the contemplation of and reflection on its verses. In offline settings this ritual manifests in different forms, for example, becoming acquainted with the verse's meaning, its lessons, and its guidance. Muslims are encouraged to perform it alone or in groups, during holidays or during the 5-times prayers for the purpose of staying connected to God according to the Quran verse "A blessed Book that We have sent down upon thee, that they may \textit{contemplate} His signs and that those possessed of intellect \textit{may reflect}." (38:29) and "When the Qur'an is read, listen to it with attention, and hold your peace: that ye may receive Mercy" (7:204).

 Reminding people of this ritual appeared as another devotional act that motivates people to sharing Quran on Twitter (Dhikr in Arabic). 68 survey respondents indicated that reflecting on the meaning of the verse and reminding people of that meaning is what motivates them share Quran verses. A survey respondent (SR-29) said, "[sharing Quran on Twitter] spreads God word and reminds people of what is important, like God said in the verse (And remind, for truly the Reminder benefits the believers)" (51:56). People share verses that reflect a closer meaning to them or to an event they are going through (e.g., a family sickness, loss, or work hardship).  In the interview with (M-1), who rarely shared Quran verses on Twitter,"I worry about the behavior of tweeting Quran verses on Twitter always because I don't want the verse to be overshared or used and loses its meaning. For me the exception is when I'm going through tough times with an ill parent or friend and look towards Twitter to seek relief by sharing with friends." Another respondent said that sharing verses is a "\textit{reminder of the Quranic wisdom that can be applied in practical life}" (SR-11).

  On Twitter, sharing these verses took different forms, either through the text body of the tweet or by sharing a decorated image that includes the verses. They are also shared depending on the context of the tweet for either their literal or metaphorical meaning. This behavior is well supported by the nature of the top shared verses and topics as they mostly center around mercy, forgiveness, reward for patience, and the Prophet morals reflecting the purpose of sharing is for personal reflection and contemplation and to remind others of this ritual. (F-1) illustrates this in her comment when asked what Quran verses does she share, "Sometimes the issue is personal and I don't want to share it with everyone, but I want them to think of me and pray for me. So I choose short verses that are used to indicate the hardship and hope, without needing to share the details. It give me hope when I see people responding with Amen or other wishes". In talking about her approach to sharing Quran on Twitter, F-1's preference is to maintain privacy on the personal issue, however, share abstract knowledge to gain other's support. She does that through sharing Quran verses that hold a metaphorical meaning to her hardship signaling to others, who understand the verse meaning, her current circumstances. By sharing verses on hope and hardship, she is practicing personal reflection while also seeking external support from her followers without giving much of the personal details through the known meaning of the verse. In context of Islam, sharing religious content for grieving and allowing others to comprehend ones agony  has been discussed in the form of photo sharing \cite{alshehri2018beauty}. Building on the observation made in Alshehri's study \cite{alshehri2018beauty}, we contribute an additional form of expression through Quran verses with the purpose of reflection and contemplation for one self and those who follow them.

 This theme also helps explain what we observed earlier regarding the difference is topic coverage by different account types and the most shared themes with verses. For example, personal accounts post more Quran verses focused on \textit{Hereafter and Unseens} and \textit{Worship} than other categories.

 \subsection{Expression of solidarity}
 Expressing solidarity, showing compassion, and altruism are central values in Islam explained in the following hadith: "\textit{Help them [your neighbor] if they ask for your help, give them relief if they seek your relief, lend them if they need a loan, show them concern if they are distressed, nurse them when they are ill, attend their funeral when they die, congratulate them if they meet any good, (...)}"~\cite{bukhari1981sahih}. The use of the term "neighbor" extends beyond the physical neighbor to include people in the wider community as well as Muslims throughout the world. 

 Our analysis revealed that Quran verses are shared on Twitter to express solidarity during global events and crises, whether against Muslims or not. From the survey, 11 respondent indicated they share Quran verses as an expression of solidarity and support. A survey respondent said, "\textit{Quran verses hold meaning that give hope and is relevant to cite for current political or social events} (SR-115). Another said, "\textit{[I share Quran verses] to spread a meaning or show my support to a situation}". Arab-Muslim Twitter reflect on how they find Quran verses that offer solace and hope meaningful to share during global crises and events.
 
 Almost all interview participates discussed showing support during global crises as a reason for either tweeting or retweeting Quran verses. (M-3) reflects, "during events like the Arab Spring or mosque shoots, I felt lost of words, what can one say?! So I like or retweet many of the Quran verses shared on the event hashtag to amplify my support." In his reflection, M-3 clarifies how Quran verses helps him express his emotions of sadness and support online better than his own words. Using different social media outlets during tragic events to express solidarity with others is becoming a typical practice and part of how people converge online and collectively deal with events. Expressing condolence, solidarity, and support on social media using drawing, hashtags, slogans, and images following terrorist attacks is becoming a common practice \cite{lin2014ripple, cvetojevic2018analyzing, de2018expressing}. Many of these slogans and hashtags are extending beyond one event to becoming general marker of solidarity and alignment to other attacks \cite{de2018expressing}. Similarly, in our study, we found that specific Quran verses are becoming symbols for specific types of events, like aggression towards or attacks against Muslims serving the same purpose of providing hope, solace, and consolation.

 \subsubsection{Christchurch Mosque Event}

To further our analysis of this theme, we collected Arabic tweets for three
days after the tragic terrorist attack on the Christchurch mosque in New Zealand in March 2019 took the lives of 50 Muslims during prayer~\cite{newzealand_attack}. We used the Arabic hashtag for the word New Zealand, and collected tweets for the three days that followed the attack, we then extracted all the tweets that shared Quran verses using our algorithm. We found around 295K tweets, out of which 212K containing Quran verses or fragments. This significant percentage indicates that Arab Muslims used Quran verses to express solidarity and sadness on Twitter. Looking closely at our data set for the top shared complete verses in our sample reveals that the two Quran verses that were mostly shared with over 9K and 5.5K shares respectively, were \textit{"They desire to extinguish the Light of God with their mouths, but God completes His Light, though the disbelievers be averse"} (61:8), referring to the attacker, and \textit{"For what sin she was killed"} (81:9), which is a metaphorical use that references the victims. In the former verse, the "Light of God", is interpreted as a reference to Islam ~\cite{nasr_2017} and is in this context is especially meaningful because the location of the shooting event was during prayer time inside a mosque. 

Verse fragments occurred even more in the tweets; the top shared fragment verses with more than 55K shares was \textit{"But it increased them in faith, and they said, 'God suffices us, an excellent Guardian is He!'"} (3:173). This verse means that those who have been wronged should not be discouraged when others unite to harm them; it is a verse of encouragement and strength, a sentiment which resonated with thousands of Arab Muslim Twitter users in the shooting aftermath. 

\subsection{Automating Quran Sharing on Twitter}
 Automating Quran sharing through specialized apps appeared in both the quantitative and qualitative analysis. From our qualitative analysis, 25\% survey respondents indicated they use apps to tweet Quran verses and other religious content on their behalf, while almost all interview participants indicated they have used or encountered these app on their feed. The mere fact that these applications exist and are highly used, indicates that this behavior is common and supported. take, for example, du3a.org (@Du3a\_Org) which has more than 159K followers and requires users to sign up with their Twitter account, and authorize it to tweet Quran verses on their behalf in the frequency they set. Tweets from these applications are indicated with a source indicator in the tweet body, for example, "this tweet was generated by @Du3a". Motivating the use of these applications is the ritual of \textit{ceaseless charity}; by sharing religious content on ones timeline, without the need to post it manually, they ensure a consistent stream of deeds ~\cite{kauthar_2017}.

  Commenting on these application, an interview participant, (M-2) said, "I personally don't see value in using these apps, because there is no logic to the machine gaining the deeds on behalf of the human. However, I know people who use it as a source of ceaseless charity. I feel more education on this trend is required". M-2's reasoning with the app helps explain the trend we found in the earlier quantitative analysis on the higher percentage of retweets for manually generated Quran tweets compared to those generated by apps.
  Conversely, (F-4) commented, "When I was traveling abroad to study and was sure I was going to be away from my Twitter account, I signed up to an app that tweets Quran verses for me. It felt better than leaving my account vacant...at least I get the rewards". For F-4, her desire to continue the stream of good deeds is supported by the app functionality to automate tweeting, which she perceives more valuable than keeping her Twitter account vacant. Leveraging automation to support religious practice and ritual is common amongst the Orthodox Jewish community as they observe the Sabbath and other holidays\cite{woodruff2007sabbath} to lift the burden of performing mundane activities and offer more focus spiritual time. In F-4's case, automation is leveraged to perform the ritual itself, i.e. ceaseless charity, offering a new perspective on how some Arab-Muslims conceptualize automation and utilize it for religious support.

 The opposing opinions to the use of these apps also appeared in the qualitative analysis with 18 survey respondents and 3 interview participants sharing concerns and referring to the misconception that needs to be further discussed and tackled by Muslim scholars. One respondent indicated, "people believe that by simply posting a Quran verse on their account that they will receive the good deed without any effort just by people reading them. They also believe it works as a ceaseless charity for them after death!" (SR-37). Similarly F-4 said, "In my opinion it has no benefit and it is just saying God's name in vain. But some think it is of benefit to others and quoted the verse "\{And remind, for truly the Reminder benefits the believers.\}". While some see no value in these automatically generated tweets, others considered them a click bait of some sort. Illustrated in F-1 comment, "I know a friend that uses these apps to tweet Quran, but she is not that religious. In my opinion she is doing it to to gain a religious favour with people and as a result gain more Twitter followers. Its a fruitful trend gain more followers or retweets". F-1 cites that using these apps to share religious content  not only granted them "religious favor", but also reflected positively on their image amongst their followers. As with previous examples, this excerpt underscores the conflict that can occur as Arab Muslim users negotiate the transition of offline rituals to online space, dealing with the intention and the audience perception.

\subsubsection{Digital memorial accounts.} 
 Offline, Muslims are encouraged to memorialize loved ones by reciting Quran in their honor, dedicating dua for them, and creating sources of charity on their behalf.  During the manual labeling of account types (section 5.2), we observed a number of accounts that were dedicated for deceased users. Looking closely at these accounts, they were registered and maintained Twitter profiles by closer ken of the deceased who register their personal account [the ken's account] on applications like Du3a.org that will take on the responsibility of tweeting Quran verses regularly on behalf of the deceased with the goal of helping them continue the stream of charity. We refer to these profiles as \textit{digital memorials}. A few participants discussed the nature and purpose of these accounts and whether they are valuable or not and to whom.   F-1 reflected on a recent loss where her cousins established an account for her uncle (their father) on Twitter to share Quran verses and dua. When I asked about the value of these accounts she said, "they are great reminders for them and us to continue to pray for him. So if I encounter his account on Twitter, I would right away think of him and say a prayer. Reading the dua shared by the application helps in the moment". In her answer, F-1 is clear about the purpose of these accounts is to serve as reminders for the viewer to supplicate and share a prayer for the deceased rather than the app doing that for them. In addition, she explains how the dua or Quran verse associated with the tweet helps to serve as a memory aid for what to pray.
 
   While F-1's answer reflects the value of having these apps and accounts, M-2 had a different opinion. He brought up these accounts in the same context of automation, which he disapproves, as an example of "complacency on behalf of the deceased family". He said, "this is a wrong interpretation of ceaseless charity for the deceased. There are many ways offline to ensure the deceased stream of good deeds continues that do not include Twitter." While digital memorial are becoming a wide practice, M-2's answer reflects that questions regarding the viability of this modern approach and whether it is really doing due diligence for the deceased. 
 
   Overall, this practice sheds a new light on the value of Twitter as a platform to maintain familial relationships, whether alive or dead. This discovery adds a new dimension, a Muslim one, to Brubaker, Hayes, \& Dourish's ~\cite{brubaker2013beyond} study on how bereaved users reappropriate Facebook for memorializing the dead where we also seek answers to what happens if Twitter algorithms detects, wrongfully so, the automated pattern on these accounts and decided to deactivate them? Underscoring the need for additional policy and design considerations that respectfully and consciously consider the dead and their grieving families.

\section{Discussion}
 Religion and technology are intertwined forces, yet most scholarly venues are established in a way lent to their separate study. In our study, we sought to bring these forces closer through an exploration of how Arab-Muslims express and practice Quran-related rituals on Twitter. By quantitatively analyzing a Twitter dataset of 2.6 million Arabic tweets sharing Quran that have been retweeted over 9 million times, followed by a qualitative analysis of survey responses (n=135) and seven in-depth interviews, we identified ways in which Arab-Muslims enact religious values online and the affordances of social media that support them.
 
  In line with previous research on how ``the internet is being studied and conceived of as a sacramental space'' where religious users present their beliefs and practices \cite{campbell2005considering}, we extend this idea to the understudied context of the Arab Muslim world to gain a better understanding of the way Islamic cultural values impact how Arab-Muslim social media users present themselves online. We suggest that Twitter affordances are reshaping what it means to be religious in society and as a result reshaping offline religiosity. We suggest that sharing Quran verses on Twitter, whether manually composing tweets or via apps, supports the enactment of religious rituals that often practiced offline. We provide examples of these rituals like performing ceaseless charity, creating digital memorials, practicing contemplation and reflection, and expressing solidarity with other Muslims. None of these are new Islamic rituals; however, they are rituals that are re-envisioned by users' creativity through the affordances of technology.
 
 One aspect of our analysis was to compare the topic distribution of the Quran to the topic distribution of verses shared in our dataset. We found that among the thirteen categories of Quran, the topic distribution of Quran verses that are shared on Twitter does not match the distribution of topics in the Quran (see Figure 2). Verses on Sharia law, worshipping God, and stories of Prophet Muhammad are covered in the Quran much less than the Twitter distribution would suggest (represents over 50\% of the share volume on Twitter). Whereas verses on the Hereafter and Unseens, and Stories of Prophets are amongst the most popular Quran topics but not so amongst Twitter users. Twitter, as a global and social platform, is conceptualized by this population as a a centralized space for knowledge sharing, seeking, and archiving, which makes Muslims from around the world gravitate to it to learn from the scholars they follow and admire. As we see in the data (Table 3), these religious scholars are the ones with the largest number of retweets on verses that include instructions and guidance on Islamic practices. 

 Our study findings build on earlier research that shows how Islamic rituals and values are being re-envisioned on social media platforms, such as Muslim marriages~\cite{al2017against}, Islamic privacy~\cite{abokhodair2017transnational}, and Tatbeer ~\cite{alshehri2018beauty} to demonstrate the remarkable adaptability and ingenuity of this user group as they embrace these global technologies and use them in ways that allow them to maintain the norms and values important to their families, religious beliefs, and communities. As we have shown, when users engage in acts of worship through Quran sharing on Twitter, they do so with focus on sharing verses on topics such as the hereafter, God's mercy, and sharia law, and far less about topics like Jihad and recruitment.

  In fact, our analysis shows that Jihad is not discussed frequently in the Quran and is similarly not often shared on Twitter (making up only 6.5\% of the share volume in Figure 2). Looking closely at the usage of the verses shared on this topic, our analysis found that the top shared verses in this category are not referring to the Jihad that is focused on fighting the enemies of Islam, but rather to the "greater Jihad" (Ijtihad is Arabic) that means the spiritual struggle within oneself against sin.  These findings adds an additional dimension to findings from earlier research that studied Quran sharing on Twitter for the purposes of understanding Jihad and extremist groups recruitment online where these Quran verses suffered the extreme religious interpretations by terrorists \cite{venkatraman2007religious, mueller_rashbaum_baker_2017,trump_2016} showing that sharing Quran verses goes well beyond the miscontextualized reporting. Moreover, these studies' focus on extremist content at the without adopting a wider lens misrepresents this activity as representative of online Islamic life with damaging consequences for the Muslim community.

 Only through reading social media activity through a culturally situated lens does its situated meaning becomes clear. For instance, the prominent Islamic belief in the afterlife manifested in what users shared online and how they shared it (manually or through apps). The majority of the people tweeting Quran verses as a ceaseless charity or for contemplation and reflection believed that these Quran tweets will outlive them and continue granting them good deeds after their deaths. These rituals were further extended through the use automation, working with the belief that they will accumulate good deeds every time someone reads the tweet, engages with it, or retweets it. In line with \cite{abokhodair2017transnational}, Arab-Muslim users treat their social media profiles as permanent extensions of themselves: what applies to these profiles and how they are perceived by others extends to their creators. In considering the entirety of a user's contemporary life experience and faith, and continuing to understand the bi-directional effect of technology on their spiritual practice, we are more able to communicate a complete picture of their lives.
 
  We believe that modern technology will increasingly subtend new forms of religious and spiritual experiences and continue to influence offline rituals, similar to the projections made by the Pew Research Centered in 2004 on the internet~\cite{hoover_clark_rainie_2004}. Therefore, in our globalized world, further research needs to probe the interaction between religious and secular orientations on the way social media users from more traditional societies adopt and adapt applications designed outside those local contexts. Instead of assuming a universal secular identity position for social media users premised upon one set of cultural assumptions where religious association is less valued, we urge the CSCW community to consider the religious dimension of online expression for certain populations of users and take it into account when studying users and designing culturally relevant applications.

\subsection{Policy and Design Considerations} 
  Studying Islamic rituals online and learning the importance of our users' religious forms of online expression inspires desiderata for the design of culturally relevant applications for users like those represented by our participants. 
 \subsubsection {Social media profiles and death policies} When taking the digital memorial accounts for example, we find the Twitter's \textit{after death} policy\footnote{https://help.twitter.com/en/rules-and-policies/contact-twitter-about-a-deceased-family-members-account} lacks a deeper understanding of this practice from the Muslim perspective. The policy offers two options: deactivating the account by requiring an immediate family member to provide proof of death, or leaving the account as-is, which leaves it open to spam bot attacks\cite{abc_news_2016}. Our study results surface the need to further enhance these capabilities through design and policy. One potential enhancement could be the introduction of features supporting activities considered to be acts of ceaseless charity as part of memorialized accounts. This could be done by identifying the Quran verses a user had posted on their account and resurfacing these periodically. Another enhancement could focus on extending the role of the ``legacy contact'' currently available on Facebook and Instagram, where users can assign a specific person to administer their page in the event of their death to include turning the deceased account into a ``memorialized account''\footnote{https://www.facebook.com/help/1506822589577997/}.
 
 \subsubsection{Automation policy risks} Our participants expressed the importance of these digital memorial accounts in giving solace and providing comfort to the grieving families of the deceased. Our findings show that majority of these accounts are managed through applications that automate the sharing of Quran verses and dua. Relatedly, a concern with these accounts is that they get suspended by the platform automation policy. Therefore, better detection is required for differentiating these accounts from bots or fake accounts, so they are not inadvertently shut down. 

 \subsubsection{Social media as everlasting platforms} Our analysis shows how this user population might conceive these platforms as everlasting in the creation of digital memorial and automated ceaseless charity accounts. A relevant policy consideration is concerned with putting policies in place to carefully handle these accounts in case these social media platforms go out of business. One suggestion is to inform the ``legacy contact'' of the discontinuation of the platform and offer them the option to transfer the account to a different service, or to download the content. 
 
 We believe that reconciling technology with religion will expand the horizon for technology design and its application beyond what was imagined while simultaneously enhancing users' experience with technology in ways we have not considered before.

\subsection{Limitations}
 Although our study has provided some new insights about Quran sharing on Twitter and has answered our research questions, we acknowledge some limitations. With regard to our data collection approach, we note a few: there are inherent constraints in using the Twitter streaming API as a data source,  where it only allows retrieving tweets with specific limits, hence possibly missing part of the full stream of tweets containing our search keywords. In addition, there is the restriction of phrases we used for sampling, which limits the data we analyzed. Relatedly, our collection method focused on achieving high precision, as we looked for exact matches of verses shared in tweets to our references in the Quran with a minimum three-word match, thus we missed tweeted verses that have misspelling and verses of fewer than three words, which are common in the Quran\footnote{ \url{https://www.qurananalysis.com/analysis/basic-statistics.php?lang=EN}}. Unfortunately, using fewer than three words caused an unacceptable level of noise in the results. In addition, our data collection did not consider the Quran verses in images, which we anticipate would be a good addition to the results of this study. 

 In data labeling we also were challenged by what we suspected to be bot accounts due to insufficient validation, although we have tried to identify them when possible. In our initial experiments, we used Botometer\footnote{\url{https://botometer.iuni.iu.edu/}}~\cite{davis2016botornot} with the accounts we manually annotated to validate the tool's accuracy in detecting bots. We noticed that the tool was returning random results, which might be due to performance issues in Arabic. We replaced this with the filtering process to the application-posted tweets to solve for part of the problem and to still support the finding regarding employing automation for religious sharing. Nevertheless, we cannot claim that our \textit{human\_tweets} dataset is fully clean of tweets by bot accounts.

\section{Conclusion and Future Work} 
  In this paper, we presented what we believe to be the first exploratory study of Quran sharing on Twitter by using a mixed method approach to collect and analyze a Twitter dataset (2.6 million tweets that contained the Quran) complemented with a survey study (n=135) of Twitter users, among which we interviewed seven, to understand why people share the Quran on Twitter. We found that Arab Muslims share the Quran on Twitter as a form of re-envisioning Islamic rituals, such as the ritual of ceaseless charity, remembrance agents for deceased loved ones, and as a common language to express solidarity with other Muslims during hard times. The study shows the effect of and exchange between technology, Muslim society, and the Islamic faith ~\cite{pinch1984social}. Lastly, we observe automation practices on Twitter to support some Islamic rituals and find that Arabic Twitter users register applications to post Quran verses on their behalf as a form of religious devotion, although that content is less engaging compared to content posted manually by users themselves. 
  
  For future work, a more advanced Quran-verse-detector (Quran Classifier) could be built to be able to identify short (less than 3 words) verses, long verses, and image-based Quran verses in tweets with high precision and recall. In addition, deeper qualitative research is necessary to further unpack the findings of this work, and would benefit from the further use of qualitative methods to probe underlying meanings, for instance a co-design session with users to imagine techno-spiritual technologies which would be more sensitive to the needs of religious and spiritual users. In our survey, some of the users provided contact information for additional future interviews. This is one direction we aim to explore as future work, in order to learn more about the practice and opinions of people on Quran sharing, and the expected role of technology in this domain. Continuing this line of research will help further the discussion around religion and spirituality in the context of computer-mediated communication studies beyond our focus on Islam populations.

-------------------------------------------------------

\bibliographystyle{ACM-Reference-Format}
\bibliography{bibfile}

\newpage
\appendix
\section{Quran Verses in Arabic}

\begin{center}
\begin{table}[H]
\footnotesize
\begin{tabular}{l p{12.5cm}} 
\hline
\textbf{Verse} & \textbf{Original Arabic Verse}\\
\hline
(48:1) & \<إِنَّا فَتَحْنَا لَكَ فَتْحًا مُبِينًا>\\
(4:110) & \<وَمَنْ يَعْمَلْ سُوءًا أَوْ يَظْلِمْ نَفْسَهُ ثُمَّ يَسْتَغْفِرِ اللَّهَ يَجِدِ اللَّهَ غَفُورًا رَحِيمًا>\\
(33:56) & \<إِنَّ اللَّهَ وَمَلَائِكَتَهُ يُصَلُّونَ عَلَى النَّبِيِّ ۚ يَا أَيُّهَا الَّذِينَ آمَنُوا صَلُّوا عَلَيْهِ وَسَلِّمُوا تَسْلِيمًا>\\
(68:4) & \<وَإِنَّكَ لَعَلَىٰ خُلُقٍ عَظِيمٍ>\\
(113:1) & \<قُلْ أَعُوذُ بِرَبِّ الْفَلَقِ>\\
(113:5) & \<وَمِنْ شَرِّ حَاسِدٍ إِذَا حَسَدَ>\\
(113:2) & \<مِنْ شَرِّ مَا خَلَقَ>\\
(113:3) & \<وَمِنْ شَرِّ غَاسِقٍ إِذَا وَقَبَ>\\
(112:1) & \<قُلْ هُوَ اللَّهُ أَحَدٌ>\\
(112:3) & \<لَمْ يَلِدْ وَلَمْ يُولَدْ>\\
\hline
\end{tabular}
\caption{Top 10 shared Quran complete verses in \textit{human\_tweets} collection in Arabic.}
\label{tb:topCompletear}

\begin{tabular}{l p{12.5cm}} 
\hline
\textbf{Verse} & \textbf{Original Arabic Verse}\\
\hline
(65:1) & \< لَا تَدْرِي لَعَلَّ اللَّهَ يُحْدِثُ بَعْدَ ذَٰلِكَ أَمْرًا
>\\
(19:64) & \<وَمَا كَانَ رَبُّكَ نَسِيًّا>\\
(8:33) & \<وَمَا كَانَ اللَّهُ مُعَذِّبَهُمْ وَهُمْ يَسْتَغْفِرُونَ>\\
(26:227) & \<وَسَيَعْلَمُ الَّذِينَ ظَلَمُوا أَيَّ مُنْقَلَبٍ يَنْقَلِبُونَ>\\
(40:21) & \< فَأَخَذَهُمُ اللَّهُ بِذُنُوبِهِمْ وَمَا كَانَ لَهُمْ مِنَ اللَّهِ مِنْ وَاقٍ>\\
(7:100) & \< أَنْ لَوْ نَشَاءُ أَصَبْنَاهُمْ بِذُنُوبِهِمْ ۚ وَنَطْبَعُ عَلَىٰ قُلُوبِهِمْ فَهُمْ لَا يَسْمَعُونَ>\\
(19:21) & \<قَالَ كَذَٰلِكِ قَالَ رَبُّكِ هُوَ عَلَيَّ هَيِّنٌ ۖ >\\
(5:32) & \<وَمَنْ أَحْيَاهَا فَكَأَنَّمَا أَحْيَا النَّاسَ جَمِيعًا>\\
(7:156) & \<وَرَحْمَتِي وَسِعَتْ كُلَّ شَيْءٍ ۚ>\\
(14:7) & \<لَئِنْ شَكَرْتُمْ لَأَزِيدَنَّكُمْ ۖ>\\
\hline
\end{tabular}
\caption{Top 10 shared Quran verse fragments in \textit{human\_tweets} collection}
\label{tb:topIncompletear}
\end{table}
\end{center}

\newpage
\section{Survey Instrument (Translated to English)}
As-salamu alaykum
We are a research team from the universities of Edinburgh and Washington studying the sharing of Quran verses on Twitter. Our goal at this stage is to understand the reasons behind this behavior. Help us by answering the following questions honestly. Thank you!

\begin{itemize}
    \item Gender: (Male, Female, Prefer not to declare) 
    \item Age (text field)
    \item How many times do you use Twitter in a week for tweeting? (daily, 3-4 times a week, 1-2 days a week, I don't use Twitter for tweeting, other)
    \item How many times do you use Twitter in a week for reading others tweets? (daily, 3-4 times a week, 1-2 days a week, I don't use Twitter for reading, other)
    \item Do you follow accounts that tweet Quran verses or encounter them on Twitter? (yes, no, other)
    \item Do you share Quran verses on Twitter? (yes, no, other)
    \item If yes, how often do you share Quran verses on Twitter? (1 = only 1 time - 5 = Always scale answer)
    \item How do you tweet Quran verses? (always manually composed, always app generated, hybrid - sometimes manually composed and sometimes app generated, Other)
    \item (if answer is app generated or hybrid) Please tell us why you use apps to generate Quran verses? (open-ended)
    \item Why do you share Quran verses on Twitter? (open-ended)
    \item In your opinion, what is the purpose for sharing Quran verses on Twitter? (open-ended)
    \item We would love to hear more from you, are you open to speak more about Quran sharing on Twitter? (yes, no)
    \item if yes, please provide the best way to contact you (text box)
\end{itemize}

\end{document}